\documentclass[11pt]{amsart}
\usepackage[top=1.5in, bottom=1.5in, left=1.5in, right=1.5in]{geometry}
\usepackage{graphicx}
 \usepackage{amsaddr}
\usepackage{amssymb}
\usepackage{multirow}
\usepackage[sort&compress,square,comma,authoryear]{natbib}
\usepackage{times}

\newcommand{\D}[0]{\mathrm{d}}
\newcommand{\erf}[0]{\mathrm{erf}}
\newcommand{\dvt}[0]{\Delta \vartheta}
\newcommand{\svt}[0]{\sigma_{\vartheta}}

\title[Small Satellite Optical Communication Networks]{Small Satellite Optical Communication Networks: Analytical Models}
\author[Niraj K. Inamdar]{Niraj K. Inamdar}
\email{inamdar@mit.edu}

\begin{document}
\maketitle
\begin{abstract}
Small satellites, especially picosatellites, appear poised to play an important role in the future of space systems. Due to their size, however, integrating them with 
high-throughput laser-based communication systems remains a challenge. In this paper, we develop several analytical models that quantify how optical communication networks can be implemented with picosatellites. We do so with the goal of identifying design challenges that need to be addressed if picosatellites and optical communication systems are indeed to be a standard part of the future space technological landscape.
\end{abstract}

\tableofcontents

\section{Introduction}
It appears that small satellite technology, and picosatellites in particular, will be a significant part of space system architectures in coming years. Architectures reliant upon so-called monolithic (or, more descriptively, aggregated) spacecraft in which a broad range of highly-performing subsystems are brought together into a complex, consolidated system are expensive (on the order of hundreds of millions to billions of dollars) and require significant development times. Small satellites, on the other hand, with dimensions on the orders of tens of centimeters typically have power requirements far lower than aggregated satellites. Small satellites that rely on off-the-shelf components or have well-defined form factors that make them amenable to modularity offer significant financial savings, although at the cost of technical capability. For instance, a remote sensing system mounted on a CubeSat is expected to have significantly lower performance (measured in terms of, say, image stability, detector sensitivity, or on-board processing capability) than its counterpart on a Flagship-class or even a Discovery-class mission. 
 
The tradeoff between the two types of spacecraft in terms of technical capability can in part be mitigated by the idea of ``power in numbers'', in which a larger number of small satellites can be deployed and therefore make up for some of the lost signal. For instance, if the desired performance in terms of signal to noise ratio $\mathrm{SNR}$ is $\mathrm{SNR}^*$, but the actual performance is $\mathrm{SNR}_0$, then the number $N$ of deployed small satellites would be $N \sim \left(\mathrm{SNR}^*/\mathrm{SNR}_0\right)^2$.\footnote{This assumes integration times for all satellites are the same, and that the signal received has Poisson-type noise characteristics.} This strategy would of course only be practical if, for instance, the cost of a single small satellite was at most $1/N$ the cost of its aggregated counterpart. The strategy of regaining technical performance in numbers has an important corollary: disaggregated systems enable a system to be more resilient. 
 
Here, ``resilient'' is taken to mean that the marginal utility lost in a system due to a particular element being compromised in some way is such that the system is still able to retain its functionality up to a specified level. This is a natural feature of architectures in which capabilities are parallelized and disaggregated. These characteristics are obviously not desirable in every space system, but they lend themselves to missions where high throughput and persistent coverage is necessary (such as surveillance applications) and in which there are a large number of potential targets (e.g. asteroid characterization).

While small satellites (and especially picosatellites) offer many potential benefits, some of the factors that make them promising for wide-scale application in the future also make them challenging to implement. The greatest driver is their small size. If for simplicity we assume small satellites and their monolithic counterparts have a similar volumetric density, then the moment of inertia of the spacecraft scales as $d^{5}$, where $d$ is a typical length dimension of the spacecraft. A CubeSat may have a length scale of order ten times smaller than a monolithic spacecraft, requiring the control torque to be controlled to a precision $10^{-5}$ it would be on a larger spacecraft. This difficulty has led to the development of multi-stage attitude control systems for CubeSats developed for precise photometric applications. 
 
Another issue with widespread small satellite application is communication. Traditional satellite communication architectures rely on wavelengths of the same order of a small satellite's size, $\sim \mathrm{1-10~\mathrm{cm}}$. Communicating with small satellites using traditional communications systems is therefore a challenge, as the size of transmitting and receiving antennae must be of the same order of size as the signal wavelength and hence of similar size to the spacecraft itself. 
 
While strategies to mitigate this issue have been proposed \citep{babuscia2013inflatable}, this latter challenge has also led to proposals to develop optical-communication based small satellite networks. In particular, recent work has focused on developing architectures and carrying out testbed demonstrations of CubeSat optical communication networks \citep{morgan2017}. The purpose of this paper is to develop a simple analytical model for the mutual acquisition of satellites communicating by optical beacons and to propose a framework for the design of optical beacon networks comprising many elements and perhaps spanning interplanetary distances. 
	
\section{Characterizing Optical Links}

\begin{figure}[ht]
	\centering
		\includegraphics[width=1\textwidth]{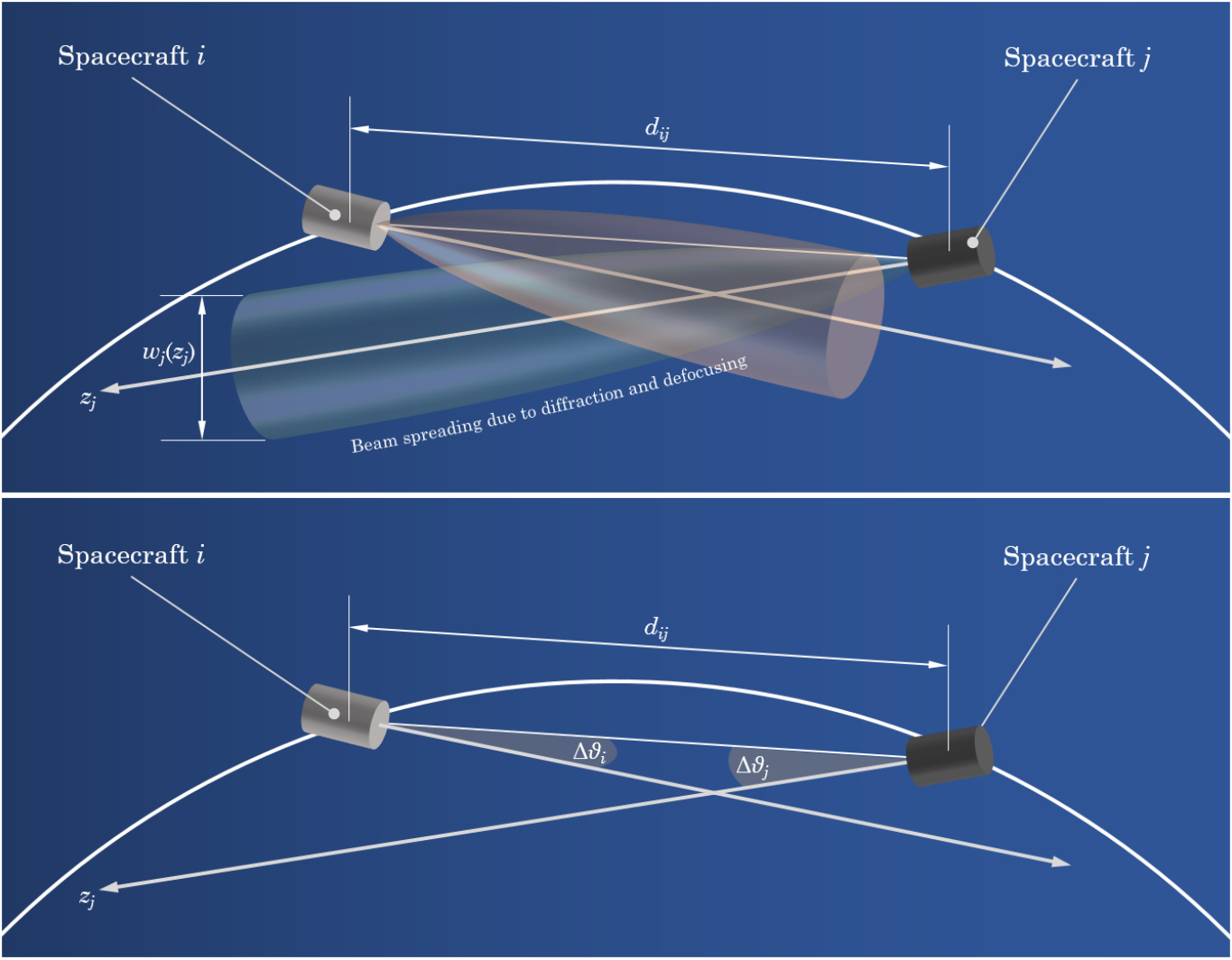}
	\caption{Overview of lasercom geometry.}
	\label{fig:Overview}
\end{figure}
For optical terminals to communicate with one another, they must first acquire one another and then maintain mutual acquisition in order to transfer data. Given picosat inertial characteristics, acquisition over large distances is challenging. In this section, we calculate the probability of acquisition based on pointing error. 

The geometry of the problem is summarized in Fig. \ref{fig:Overview}, wherein we have two spacecraft (optical 
terminals) identified by indices $i$ and $j$ and separated by a distance $d_{ij}$. 
Assume each spacecraft's optical emitter is a Gaussian beam. 
Then the power $P_{0,i}$ emitted by the $i$th spacecraft is attenuated
as a function of the mutual distance $d_{ij}$ between the two spacecraft
due to diffraction. Explicitly, the flux density profile of a monochromatic
Gaussian beam of wavelength $\lambda_i$ is given by
\begin{align}\label{eq:GBDef}
F_{i}(d_{ij}) = P_{0,i}\frac{\pi w_{0,i}^2}{d_{ij}^2\lambda_i^2}\exp\left[-\frac{\left(\Delta \vartheta_i\right)^2}{2\sigma_{\vartheta_i}^2}\right],
\end{align}
where $\Delta\vartheta_i$ is the angular displacement from the beam's center line, 
$\sigma_{\vartheta_i}$ is the spreading factor due to diffraction, and $w_{0,i}$ is the beam diameter at 
the emission point. The diffraction spreading factor is on the order of $\sim \lambda_i/w_{0,i}$; for a Gaussian
beam it is
\begin{align}
\sigma_{\vartheta_i} = \frac{\lambda_i}{2\pi w_{0,i}}.
\end{align}The beam flux can be rewritten in a more illuminating form:
\begin{align}
F_{i}(d_{ij}) = \frac{P_{0,i}}{4\pi w_{0,i}^2}\left[\frac{w_{0,i}}{w(d_{ij})}\right]^2\exp\left[-\frac{\left(\Delta \vartheta_i\right)^2}{2\sigma_{\vartheta_i}^2}\right],
\end{align}
where $w$ is the beam width as a function of distance $d_{ij}$ from the emitter. At distances more 
than $\sim w_{0,i}$ from the emitter, it is given by $w \approx \sigma_{\vartheta_i}d_{ij} = \lambda_i d_{ij}/(2\pi w_{0,i})$.
The equation above reflects the conservation of energy via the standard $\sim 1/d_{ij}^2$
rule: the prefactor $4/(\pi w_{0,i}^2)$ is the beam area at emission, and the flux is attenuated by a factor 
$(w_{0}/w)^2$ over a distance $d_{ij}$. To recover the power received, we multiply by the detector area $A_{j}$ 
at the receiver: $F_{i}(d_{ij})A_{j}$. 

\subsection{The Optical Beacon Link Budget}
The total power received can be written in the following form (see also \citet{marshall1986received}):
\begin{align*}\label{eq:PowReceived}
P_{\mathrm{Rx}} 	&= P_0   \left( \frac{\lambda^2}{4\pi d_{ij}^2} \right) \left( \frac{\pi^2 w_0^2}{\lambda^2} \right) \left( \frac{4A_{\mathrm{det}}}{\lambda^2} \right) \exp\left[-\frac{\left(\dvt\right)^2}{2\sigma_{\vartheta}^2}\right] \nonumber \\ 						
									&= P_0    L_s G_{\mathrm{Tx}} G_{\mathrm{Rx}} L_p,
\end{align*}
where 
\begin{align}
\frac{\lambda^2}{4\pi d_{ij}^2}     &\equiv L_s            ~\textrm{(space loss)},  \\
\frac{\pi^2 w_0^2}{\lambda^2}       &\equiv G_{\mathrm{Tx}}~\textrm{(Tx gain)},     \\
\frac{4A_{\mathrm{det}}}{\lambda^2} &\equiv G_{\mathrm{Rx}}~\textrm{(Rx gain)},     \\
\exp\left[-\frac{\left(\dvt\right)^2}{2\sigma_{\vartheta}^2}\right] & \equiv L_p~\textrm{(pointing loss)}.
\end{align}
The rearrangement of the terms in Eq. \eqref{eq:PowReceived} in this way gives values for the space loss, transmitter (Tx) gain,
receiver (Rx) gain, and pointing loss analogous to those used in radio communications. Eq. \eqref{eq:PowReceived}, coupled 
with an expression for signal to noise ratio ($\mathrm{SNR}$), allows us to construct the link budget. 

The $\mathrm{SNR}$ is determined by the noise characteristics of the detector, and its gain and 
quantum efficiency. Quantities of interest here 
are the photodiode gain $M$ (dimensionless), the photodiode responsivity $R_{\mathrm{PD}}$ (units of $[\mathrm{A/W}]$), the excess 
noise factor $F_{\mathrm{EN}}$ (dimensionless), and the noise equivalent bandwidth $B$ (units of $[\mathrm{Hz}]$). If $q$ is the 
fundamental charge, then the signal to noise ratio $\mathrm{SNR}$ is 
\begin{align}
\mathrm{SNR} 	&= \frac{P_{\mathrm{Rx}} R_{\mathrm{PD}}}{2 q B F_{\mathrm{EN}}} \nonumber\\
							&=  \frac{P_0    L_s G_{\mathrm{Tx}} G_{\mathrm{Rx}} L_p R_{\mathrm{PD}}}{2 q B F_{\mathrm{EN}}},
\end{align}
with the value in decibels, $\mathrm{SNR_{dB}}$, given by $10\log_{10}\mathrm{SNR}$. Typically, a floor 
is set for detection at a certain $\mathrm{SNR_{dB}}$, say $\mathrm{SNR_{dB}} \geq 3~\mathrm{dB}$. 

Note that can define the $\mathrm{SNR}$ (and
the associated threshold) in other ways. For instance, if can define a wavelength-dependent quantum efficiency of the 
detector $\mathrm{QE}_{\lambda}$ such that the number of photoelectrons generated by the detector per unit time
is $P_{\mathrm{Rx}}\mathrm{QE}_{\lambda}$. If the number of electrons generated by thermal and other sources of
noise per integration period is $\dot{n}_{\mathrm{e^{-}}}$, then we can define an $\mathrm{SNR}$
\begin{align}
\mathrm{SNR} = \frac{P_{\mathrm{Rx}}\mathrm{QE}_{\lambda}}{\dot{n}_{\mathrm{e^{-}}}}.
\end{align}
We also define for future use, a signal to noise ratio value that represents the $\mathrm{SNR}$ in the case for which there are 
no pointing losses, i.e. $L_p \equiv 1$. We denote this $\mathrm{SNR}$ with a tilde, $\tilde{\mathrm{SNR}} \equiv \mathrm{SNR}/L_p$,
and the corresponding value in decibels as $\tilde{\mathrm{SNR}}_{\mathrm{dB}}$. 

\subsection{Acquisition Probability}
Consider the case where the signal is not calculated in decibels. Terminal $i$ is seeking to acquire terminal $j$; they are at a distance $d_{ij}$ from one another. The maximum 
off-pointing angle $\dvt$ at which $j$ can still detect $i$ occurs when the pointing loss is such that $\mathrm{SNR} = \mathrm{SNR}^*$.
Solving for $\dvt$ in this case gives a maximum allowable excursion $\dvt_{\max}$ of 
\begin{align}
\left(\dvt_{\max} \right)^2 = 2\sigma_{\vartheta_i}^2\log\left(\frac{P_0    L_s G_{\mathrm{Tx}} G_{\mathrm{Rx}}\mathrm{QE}_{\lambda}}{\dot{n}_{\mathrm{e^{-}}} \mathrm{SNR}^{*}} \right)
\end{align}
Define a quantity $\tilde{\mathrm{SNR}} \equiv \mathrm{SNR}/L_p = P_0    L_s G_{\mathrm{Tx}} G_{\mathrm{Rx}}\mathrm{QE}_{\lambda}/\dot{n}_{\mathrm{e^{-}}}$. Then we can define the ratio of $\Sigma \equiv \tilde{\mathrm{SNR}}/\mathrm{SNR}^{*}$, representing the ratio of the signal to noise in the on-pointing case (with no pointing loss) to that of the detection threshold. Then we have 
\begin{align}
\dvt_{\max} = \pm\sqrt{2\sigma_{\vartheta_i}^2\log \Sigma},
\end{align}
where $\sigma_{\vartheta_i}$ parametrizes the beam width of terminal $i$.

Now suppose the angle $\dvt$ is distributed according to a probability distribution function $p_i(\dvt_i)$, which is accordingly normalized. Then the total probability of a single spacecraft being acquired is 
\begin{align}
P_i = \int_{-\dvt_{\max}}^{\dvt_{\max}}p_i(\dvt_i)\D (\Delta\vartheta)
\end{align}  
and if the misalignment of both spacecraft is independent of one another\footnote{Note that this is a conservative estimate. In 
practice, one terminal should base its acquisition strategy on } then 
\begin{align}
P_{ij} = \int_{-\dvt_{\max}}^{\dvt_{\max,i}}p_i(\dvt)\D (\dvt_i) \times \int_{-\dvt_{\max}}^{\dvt_{\max,j}}p_j(\dvt)\D (\dvt_i)
\end{align}
Suppose $p_i(\dvt)$ is Gaussian with variance $\zeta_i^2$; $\zeta_i^2$ comprises the errors in the system due to both jitter and attitude knowledge error, both control and estimation errors. These two effects would not necessarily be orthogonal to one another,
but for simplicity we may add them in quadrature so that $\zeta_i^2 = \zeta_{i,\mathrm{con}}^2 + \zeta_{i,\mathrm{kno}}^2$, where
the individual terms represents control and knowledge errors. Then 
\begin{align}
p_i(\dvt_i) = \frac{1}{\sqrt{2\pi}\zeta_i}\exp\left[-\frac{\left(\dvt \right)^2}{2\zeta_i^2}\right] 
\end{align}
and
\begin{align}
P_i = 2\int_{0}^{\dvt_{\max,i}}\frac{1}{\sqrt{2 \pi}\zeta_i}\exp\left[-\frac{\left(\dvt_i \right)^2}{2\zeta_i^2}\right] \D (\dvt_i)
\end{align}

Define the error function $\erf(x)$ using the standard form:
\begin{align}
\erf(x) = \frac{2}{\sqrt{\pi}}\int_0^{x}\exp\left(-t^2\right)\D t.
\end{align}

Then we get
\begin{align}
P_i &= \frac{2}{\sqrt{\pi}}\int_{0}^{\dvt_{\max,i}/\sqrt{2}\zeta_i} \exp\left(-t^2 \right) \D t \nonumber \\
	  &= \erf\left(\frac{\dvt_{\max,i}}{\sqrt{2}\zeta_i}\right) \nonumber \\
		&= \erf\left[\frac{\sigma_{\vartheta_i}}{\zeta_i}\left(\log\Sigma_i\right)^{1/2}\right].
\end{align}

The total probability of acquisition is 
\begin{align}
P_{ij} = \erf\left(\frac{\sigma_{\vartheta_i}}{\zeta_i}\sqrt{\log \Sigma_i} \right) \times \erf\left(\frac{\sigma_{\vartheta_j}}{\zeta_j}\sqrt{\log \Sigma_j} \right)
\end{align}
and for identical terminals
\begin{align}
P_{ij} = \erf\left(\frac{\sigma_{\vartheta}}{\zeta}\sqrt{\log \Sigma} \right)^2.
\end{align}
If we choose instead to define the signal to noise ratio metric using decibels, we rewrite $\Sigma$ as 
\begin{align}
\Sigma &= \frac{\tilde{\mathrm{SNR}}}{\tilde{\mathrm{SNR}}^*} \nonumber \\
			 &= 10^{(\tilde{\mathrm{SNR}}_{\mathrm{dB}} - \mathrm{SNR}_{\mathrm{dB}}^{*})/10}.
\end{align}


\section{Optimizing the Beam Width for Acquisition}
\subsection{The Concept of Operations}
Given an expression for $P_{ij}$, we can seek its stationary values in order to determine conditions that enable us to maximize
the acquisition probability. A realistic acquisition scheme might be the following:
\begin{enumerate}
\item Knowing the expected position of the receiving satellite from its orbit, the transmitter aligns itself towards 
		the same orientation based on star-tracker data
\item The transmitter then performs a ``search'' for the receiver using one or both of the following methods:
	\begin{enumerate}
		\item The transmitter and receiver both perform scans of the sky centered about their expected positions
		\item The beacons each defocus their beams to increase the beams' solid angle ``footprint'' and hence the likelihood of being 
			detected by the other spacecraft
	\end{enumerate}
\item Once initial, mutual acquisition is achieved, the terminals progressively fine-tune their pointing (at the same time reducing beam width for increased transmission gain) and commence data transfer.
\end{enumerate}

Once the initial acquisition has been made, regardless of whether a scanning procedure is implemented or a defocusing procedure is implemented, fine tuning in mutual pointing can be achieved through a combination of expected position, star tracker data, and detection of beam power received (which indicates how far off the transmitter is off-pointing). 

\subsection{Beam Width Modulation for Initial Acquisition}
Here, we explore the option of modulating beam width for initial acquisition based on fixed design parameters and $\sigma$. 
Beam width could be modulated randomly, or in some fixed pattern, but as we see below, there is an optimum value that depends on the
detection threshold and pointing error. We note that it is not possible to defocus the beam to arbitrarily high values, since then the transmitted flux would fall below the detectability threshold. In principle, then, focusing would be such that acquisition is marginally
achieved, from which then the beam would be progressively focused to increase gain and data rates.

To determine the optimum $\svt$ we calculate 
\begin{align}
\frac{\partial P_{ij}}{\partial \svt} = 0.
\end{align}
For simplicity, we assume both terminals are identical, so that we seek $\svt$ such that
\begin{align}
\frac{\partial}{\partial \svt}\erf\left(\frac{\sigma_{\vartheta}}{\zeta}\sqrt{\log \Sigma} \right)^2 = 0.
\end{align}
Keeping in mind that $\Sigma$ is a function of $\svt$ itself (through the $G_\mathrm{Tx}$ term), we find that the optimum
beam width for detection is such that 
\begin{align}
\log \Sigma = 1~\textrm{or}~\Sigma = \tilde{\mathrm{SNR}}/\mathrm{SNR}^* = \mathrm{e},
\end{align}
where $\mathrm{e}$ is the base of the natural logarithm. Since 
\begin{align}
\tilde{\mathrm{SNR}} &=  \frac{P_0    L_s G_{\mathrm{Tx}} G_{\mathrm{Rx}} R_{\mathrm{PD}}}{2 q B F_{\mathrm{EN}}},
\end{align}
and $G_{\mathrm{Tx}} = \pi w_0^2/\lambda^2 = 1/(4\svt^2)$, we solve for the optimum $\svt$:
\begin{align}\label{eq:OptSigAc}
\svt^* = \sqrt{\frac{P_0 L_s G_{\mathrm{Rx}}R_{\mathrm{PD}}}{4\mathrm{e}\mathrm{SNR}^{*}BF_{\mathrm{EN}}}}.
\end{align}
If our threshold acquisition $\mathrm{SNR}$ is expressed in decibels, replace $\mathrm{SNR}^{*}$ with $10^{\mathrm{SNR}^{*}_{\mathrm{dB}}/10}$.

\subsection{Active Beam Modulation}
The previous calculation for optimum beam width demodulation calculated $\svt$ by maximizing a probability distribution function
integrated over all off-pointing angles $\dvt$. Suppose now we wish to \textit{actively} change the beamwidth $\svt$. While technically more expensive, it would enable greater flexibility and have particular application to beam focusing after initial acquisition. 

Before, we assumed that the probability distribution function of acquisition was a Gaussian, with the probability 
of pointing off-axis dependent upon the pointing error $\zeta$. Now, we take a different approach. We assume 
that $\dvt$ varies as a function of time, and that the probability of acquisition at any given time 
depends monotonically on the received $\mathrm{SNR}$. Thus we need to find the $\sigma_{\vartheta} = \sigma_{\vartheta,\mathrm{def}}$ such that
\begin{align}
\frac{\partial}{d\svt}\left[\frac{1}{\svt^2}\exp\left(\frac{(\dvt)^2}{2\svt^2}\right)\right] = 0,
\end{align}
which follows directly from Eq. \eqref{eq:GBDef}. From this, we find the optimum defocusing factor $\sigma_{\vartheta,\mathrm{def}}$ to be
\begin{align}\label{eq:sigVarOpt}
\sigma_{\vartheta,\mathrm{def}} = \frac{\dvt}{\sqrt{2}}.
\end{align}

\begin{figure}[ht]
	\centering
		\includegraphics[width=1\textwidth]{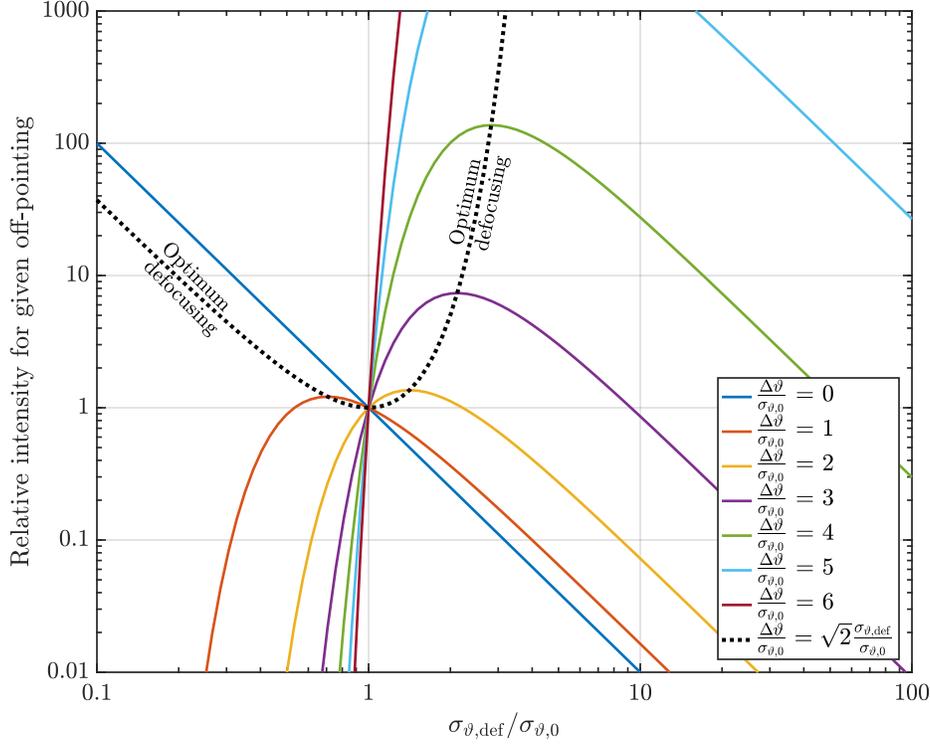}
	\caption{Optimum normalized defocusing factor based on given off-center pointing $\dvt$ (normalized to 
	initial beam width $\sigma_{\vartheta,0}$). The solid curves indicate the functional form $\exp[-(\dvt)^2/(2\svt^2)]/\svt^2$,
	which follows from the expression for a Gaussian beam for various combinations of normalized $\dvt$ and over a range of 
	$\svt$. The dotted line indicates the locus of optima found from Eq. \eqref{eq:sigVarOpt}.}
	\label{fig:DefocusFactor}
\end{figure}

\begin{figure}[ht]
	\centering
		\includegraphics[width=1\textwidth]{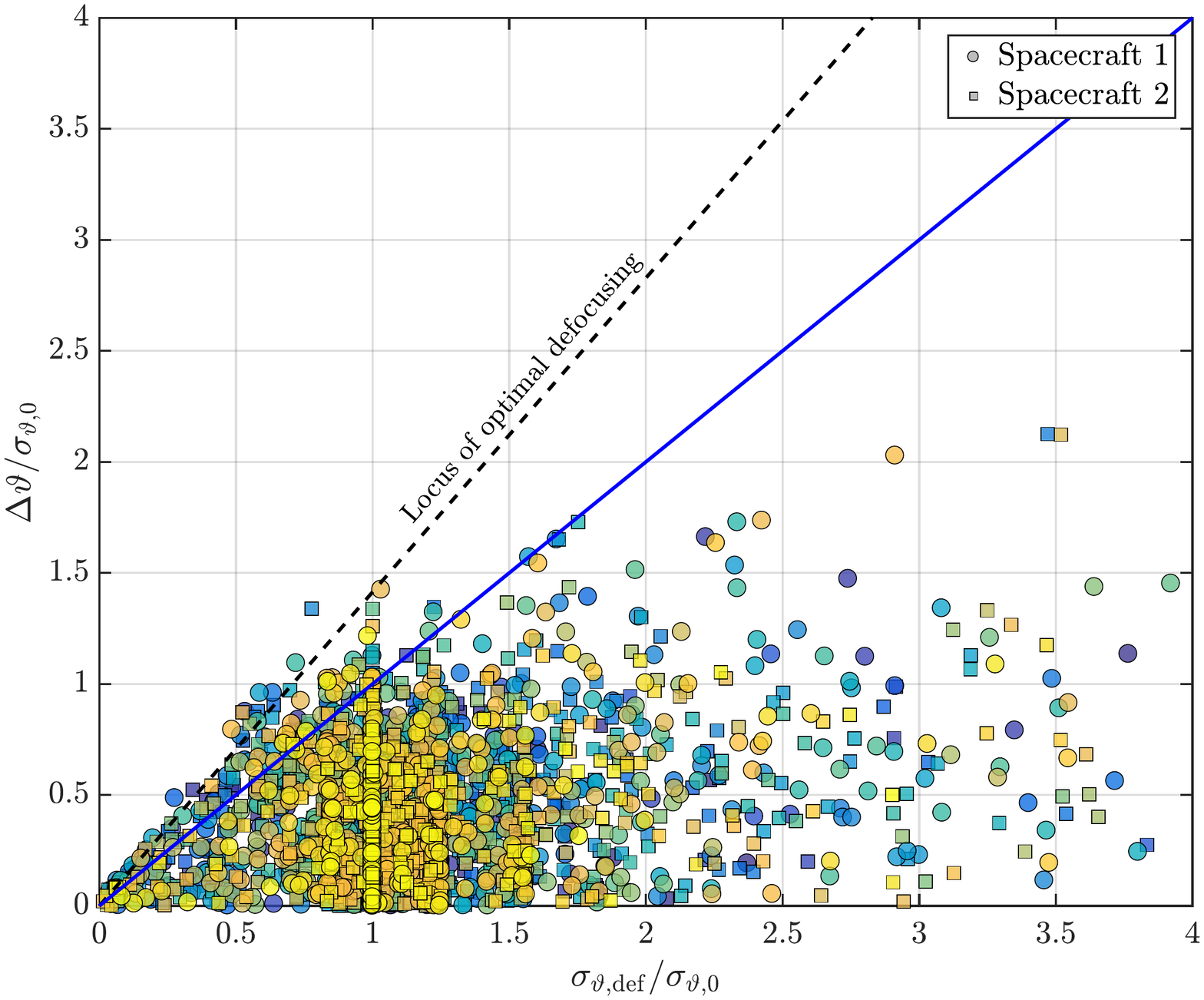}
	\caption{Optimum normalized defocusing factor based on given off-center 
	pointing $\dvt$ (normalized to initial beam width $\sigma_{\vartheta,0}$. 
	Mutual acquisitions are shown, with each individual
	acquisition indicated by color; markers differentiate spacecraft 1 and spacecraft 2. The defocusing 
	at acquisition is bounded the relationship given by Eq. \eqref{eq:sigVarOpt}. 3,000 simulations 
	were run here.}
	\label{fig:DefocusFactor2}
\end{figure}

These results are summarized in Fig. \ref{fig:DefocusFactor}, where we have normalized both $\dvt$ and $\svt$ to an arbitrary
reference value. In Fig. \ref{fig:DefocusFactor2}, we show the results of a number of Monte Carlo simulations of mutual acquisition
for identical optical terminals. In these simulations, the beam width was modulated sinusoidally, and the markers indicate the defocusing factors and off-pointing at which mutual acquisition occurred. The locus indicated by Eq. \eqref{eq:sigVarOpt} sets the practical bound at which mutual acquisition occurred in these simulations. 

\section{Developing Optical Satellite Constellations}
To demonstrate the potential practical application of the calculations above, we consider 
a simple constellation of optical terminals extending out from (say) $1~\mathrm{AU}$ to 
interplanetary distances on the order of $\sim 1~\mathrm{AU}$. The constellation comprises
the following: 
\begin{itemize}
\item A set of $n_k$ satellites on a circular heliocentric orbit at semimajor axis $a_k$ with corresponding Keplerian mean motion $\Omega_k = \sqrt{GM_{\odot}/a_k^3}$, where $G$ is Newton's constant and $M_{\odot}$ is
the mass of the Sun.
\item The satellites are equally-spaced on each ring, with an angle between each terminal of $\Delta \theta = 2\pi/n_k$ subtended from the Sun.
\item The spacing between each ring is $d_{ij}$, and the distance between closest pairs of terminals 
		on each ring is $d_{ij}$, such that $\Delta \theta = \pi/\arcsin[d_{ij}/(2a_k)] \approx 2\pi a_k/d_{ij}$. 
\end{itemize}

In this constellation, we have the benefit that all satellites on the same ring are within $d_{ij}$ of one another. The next ring of terminals at $a_{k+1} = a_k + \Delta a = a_k + d_{ij}$ has $n_{k+1}$ terminals equally spaced. The total number of terminals is then given by the sum of the $n_k$ over all rings. If we take the (good) approximation that $n_k \approx 2\pi a_k/d_{ij}$, we can calculate the total number of terminals $\mathcal{N}(a_i,a_f)$ in a constellation that spans from semimajor axes $a_i$ to $a_f$. We find
\begin{align}
\mathcal{N}(a_i,a_f) &\approx \frac{2\pi a_i}{d_{ij}}\left[ \frac{a_f - a_i}{d_{ij}} + \frac{d_{ij}}{2a_i}\left(\frac{a_f - a_i}{d_{ij}}\right)\left(\frac{a_f - a_i}{d_{ij}} + 1\right) \right]\\
										 &\approx \frac{2\pi a_i}{d_{ij}}\left[ \frac{a_f - a_i}{d_{ij}} + \frac{d_{ij}}{2a_i}\left(\frac{a_f - a_i}{d_{ij}}\right)^2 \right].
\end{align} 

On the $k$th ring, there the $n_k$ satellites are indexed by $i = 1,2,...,n_k$. If the terminals in the $k$th ring have a phase offset of $\theta_{0,k}$, then the angle subtended by the $i$th satellite is 
\begin{align}
\theta_{k}^{(i)}(t) = \theta_{0,k} + \frac{2\pi}{n_k}i + \Omega_{k}t.
\end{align}

Any two terminals indexed by $i$ and $j$ on rings $k$ and $k+1$ are separated by an angle
\begin{align}
\Delta_{k,k+1} + 2\pi\left( \frac{i}{n_k} + \frac{j}{n_{k+1}}\right) + \left(\Omega_{k} - \Omega_{k+1}\right)t.
\end{align}
Solving for $t$, we get
\begin{align}
t = \frac{1}{\Omega_{k} - \Omega_{k+1}}\left[ \Delta_{k,k+1} + 2\pi\left( \frac{i}{n_k} + \frac{j}{n_{k+1}}\right)  \right],
\end{align}
and this quantity can be minimized by choosing $i$ and $j$ such that the term in the brackets is minimized. Creating links between adjacent rings
would then mean transmitting data along one ring ($k$, say), bridging to ring $k+1$, and then moving back across to the desired terminal.

\begin{figure}[ht]
	\centering
		\includegraphics[width=0.7\textwidth]{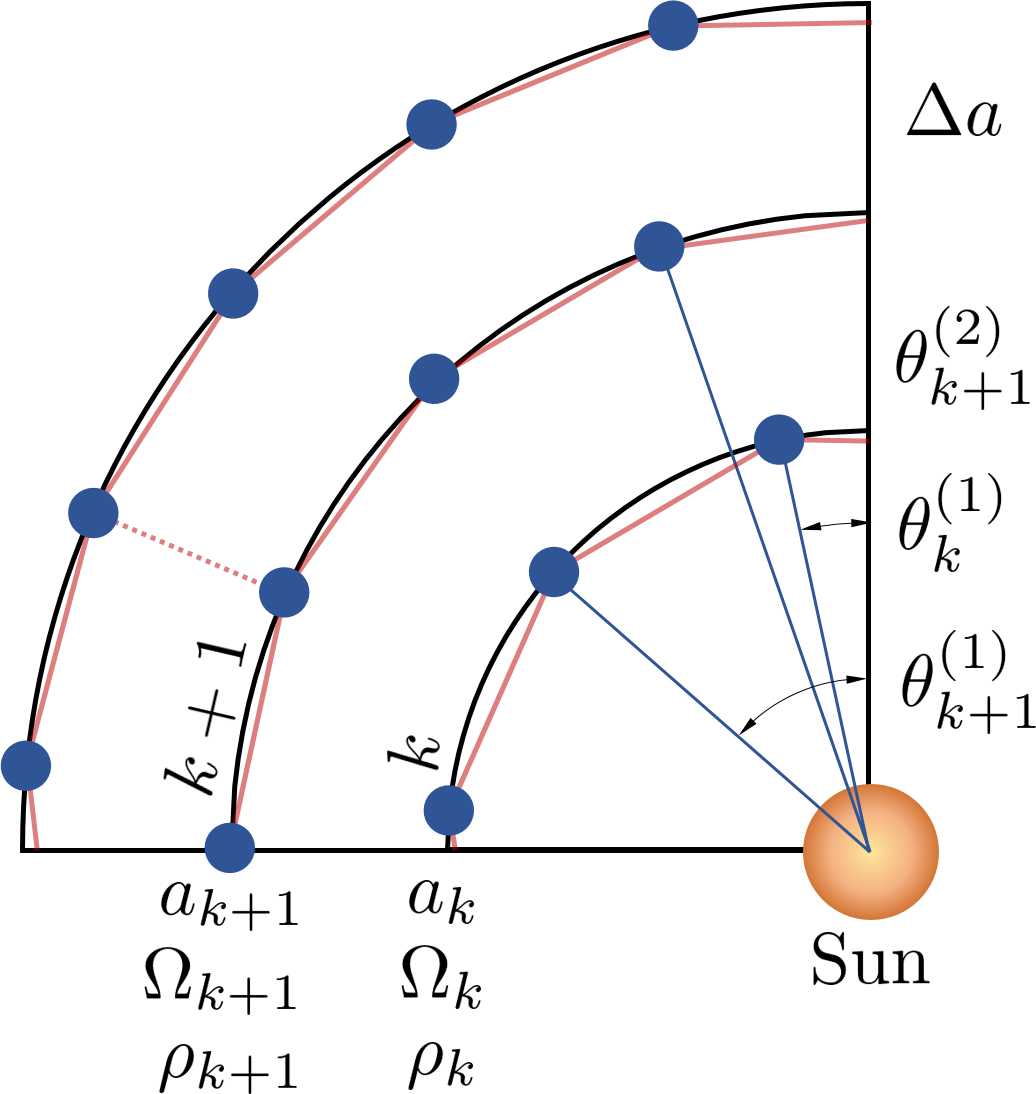}
	\caption{Constellation geometry. In this simple model, all terminals on the same semimajor axis $a_k$ have interterminal distances of $d_{ij}$, and each
						ring has a separation of $\Delta a = d_{ij}$.}
	\label{fig:Constellation}
\end{figure}

\begin{figure}[ht]
	\centering
		\includegraphics[width=1\textwidth]{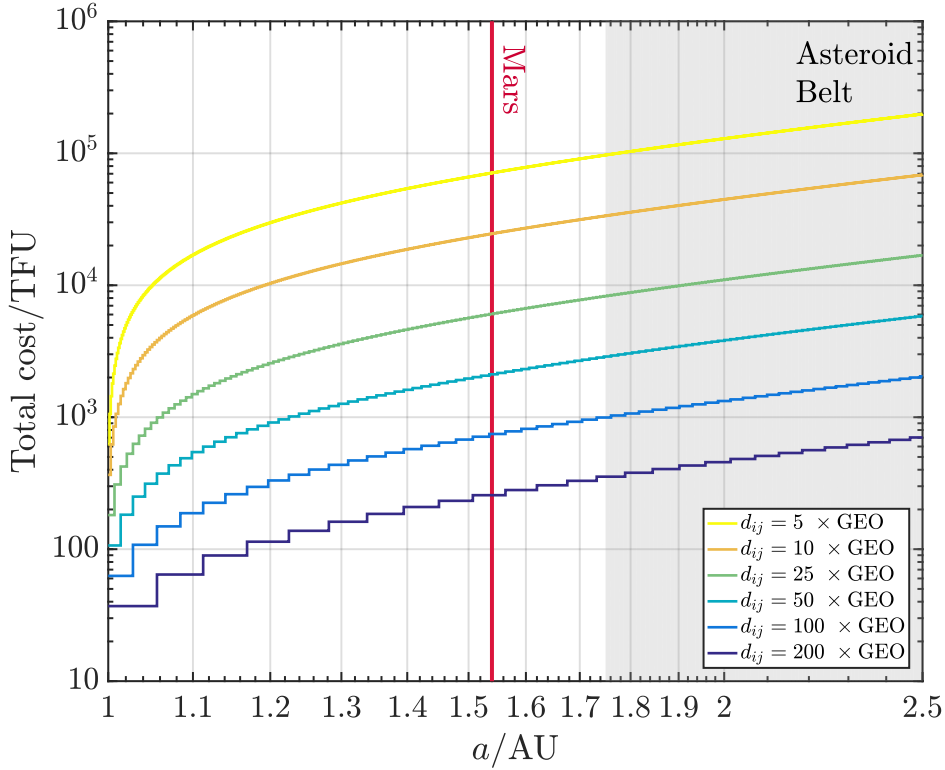}
	\caption{Constellation cost relative to $\mathrm{TFU}$ cost as a function of inter-ring spacing $d_{ij}$ and constellation extent ($a_1 = 1~\mathrm{AU}$). }
	\label{fig:ConstellationCost}
\end{figure}

The production cost is approximated by the following expression \citep{wertz1992}:
\begin{align}
\textrm{Cost} = \mathrm{TFU} \times \mathcal{N}^{B}, 
\end{align}
where $\mathrm{TFU}$ is the cost of the theoretical first unit and $B = 1 - \log[(100\%/S) - 2]$ is a factor associated with the learning curve of manufacturing. $B$ is in general a smaller number if more units are manufactured. \citet{wertz1992} recommend setting $S = 85\%$ if the number of units manufactured is greater than $50$. Since an interplanetary
constellation would require far more terminals, we set $S = 80\%$. The cost determined in this way as a function
of inter-ring spacing and constellation extent is shown in Fig. \ref{fig:ConstellationCost}.

To determine the number of terminals in the first ring, we use the following geometric criterion. 
Defining an angle $\Delta \phi$ such that $n_1\Delta \phi = 2\pi$, we should have $2a_1\sin(\Delta \phi/2) = d_{ij}$,
so that $n_1 = \lceil  \pi/\arcsin\left[d_{ij}/(2a_1)\right] \rceil$.

\section{Feasibility and Technology Frontiers}
For a $\mathrm{TFU}\sim \$ 10^{5}-10^6$, if 
an available budget is on the order of $\$ 10^{9}-10^{10}$, a value for $\textrm{Total cost}/\textrm{TFU} \sim 10^{3}-10^{5}$ suggests that with this constellation design, a network to Mars might be feasible with pointing capabilities on the order of $0.1-1~\mathrm{arcsec}$. How realistic is this? Currently, the CubeSat ASTERIA (formerly ExoplanetSat) can achieve pointing
on the order of $\sim 1~\mathrm{arcsec}$ \citep{smith2010exoplanetsat}, which it does so using a multistage attitude control system comprising a set of reaction wheels for ``coarse attitude control'', and a 
piezoelectrically-controlled focal plane that maintains fine control of the image. As mentioned above, the parameter $\zeta$ reflects error in both attitude knowledge and control. 

As it stands, based on the results 
shown in the previous section (Fig. \ref{fig:ConstellationCost}), it seems very likely that the manufacturing and cost benefits of nano- and picosatellites will lead to their forming the backbone of the network. The question we wish to ask it whether current or future technologies will enable the construction of interplanetary optical terminals. In order to do this, we first choose some baseline parameters for an optical communication terminal. The values shown in Table \ref{tab:sigBeamSumm} are taken from \citet{kingsbury2014} and \citet{morgan2017}. In Fig. \ref{fig:AcqComp_new}, we show mutual acquisition probabilities based on these hardware parameters, with varying interterminal distances $d_{ij}$ and attitude errors $\zeta$. We have assumed that the threshold for acquisition is $\mathrm{SNR}_{\mathrm{dB}}^* = 3$.

\begin{figure}[ht]
	\centering
		\includegraphics[width=1\textwidth]{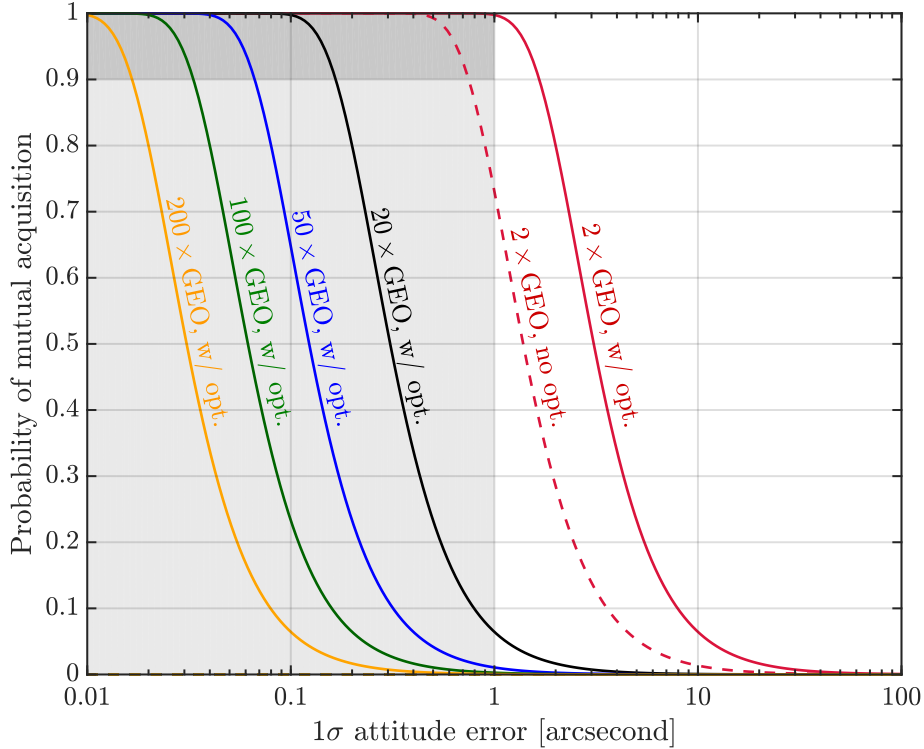}
	\caption{Probability of mutual acqusition as a function of attitude error $\zeta$ for various interterminal distances. Two cases are shown: those in which the baseline terminal parameters have been used (dotted lines)
		and those for which the beamwidth has been optimized for acquisition per Eq. \eqref{eq:OptSigAc} (solid lines). For all the cases except the closest case $d_{ij} = 2\times \mathrm{GEO}$, the probabilities before acquisition
		are practically zero. Optimizing beamwidth, however, enables significant improvement in capabilities, such that $\sim 20-100\times\mathrm{GEO}$ interterminal links are enabled. These values are to be compared to 
		Fig. \ref{fig:ConstellationCost}.}
	\label{fig:CurrFutCaps}
\end{figure}

The results in Fig. \ref{fig:AcqComp_new} have been calculated \textit{without} optimizing the beamwidth for acquisition. In Fig. \ref{fig:CurrFutCaps}, on the other hand, we show the effect of optimizing the beamwidth for acquisition. We show the $\zeta$ dependence 
of $P_{ij}$ for a range of interterminal distances between $2\times\mathrm{GEO}$, and $200\times\mathrm{GEO}$. We see that in the ``close'' case, optimizing the beamwidth can increase acquisition probability by factors of up to
8 or 9. For the ``far'' cases, we see that acquisition would not be possible with the baseline system variables---the lines corresponding to the non-optimized beamwidths are at 0. On the other hand, optimizing beamwidths 
leads to reasonable acquisition probabilities for attitude errors $\sim 0.01-1~\mathrm{arcsec}$. The ranges for which this becomes possible are to be compared with those in Fig. \ref{fig:ConstellationCost}.

\bgroup
\def\arraystretch{1}
\begin{table}
\centering
\caption{Hardware characteristics for CubeSat optical terminal \citep{kingsbury2014,morgan2017}.}
\label{tab:sigBeamSumm}
\begin{tabular}{|c|c|c|c|c|}
\hline
\textbf{Parameter name}		&		\textbf{Symbol} 			&	\textbf{Value}				& \textbf{Units} 					& \textbf{Comments} \\ \hline \hline 
\multicolumn{5}{|c|}{\textbf{Emitter}}  \\ \hline \hline 
Emitter power							& $P_0$ 									& $2.02$									& $\mathrm{W}$ 						& --- 						\\ \hline
Emitter wavelength				& $\lambda$ 							& $1.55\times 10^{-6}$	& $\mathrm{m}$  					& Nd:YAG laser			\\ \hline
Emitter aperture					& $w_0$ 									& $0.05$								& $\mathrm{m}$  					& --- 							\\ \hline \hline
\multicolumn{5}{|c|}{\textbf{Receiver}}  \\ \hline \hline
Receiver area							& $A$ 										& $(0.05)^2$							& $\mathrm{m^2}$  				& --- 						\\ \hline	
Photodiode gain						& $M$ 										& $10$									& ---						  				& --- 							\\ \hline
Photodiode responsivity		& $R_{\mathrm{PD}}$				& $0.99$								& $\mathrm{A/W}$  				& --- 							\\ \hline
Excess noise factor   		& $F_{\mathrm{EN}}$				& $4.3$									& ---							 				& --- 							\\ \hline
Noise equivalent bandwidth& $B$											& $300\times 10^{6}$		& $\mathrm{Hz}$		 				& --- 							\\ \hline
\end{tabular}
\end{table}
\egroup

\begin{figure}[ht]
	\centering
		\includegraphics[width=1\textwidth]{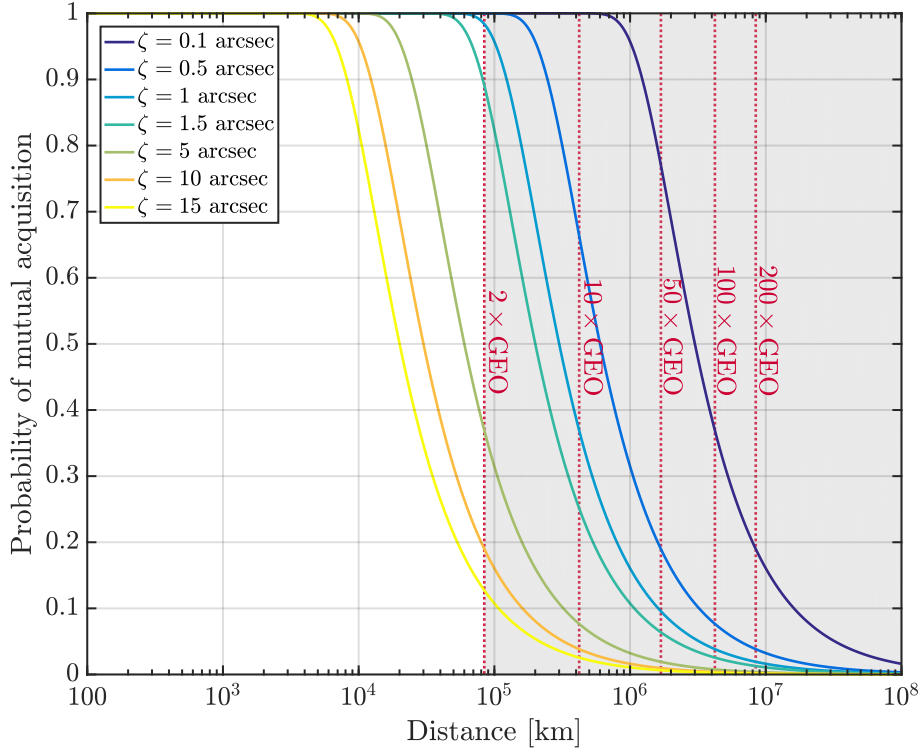}
	\caption{Mutual acquisition probabilities as a function of interterminal distance and for varying attitude errors $\zeta$. Hardware parameters are summarized in Table \ref{tab:sigBeamSumm}, and 
	the threshold for acquisition is $\mathrm{SNR}_{\mathrm{dB}}^* = 3$.}
	\label{fig:AcqComp_new}
\end{figure} 

\begin{figure}[ht]
	\centering
		\includegraphics[width=1\textwidth]{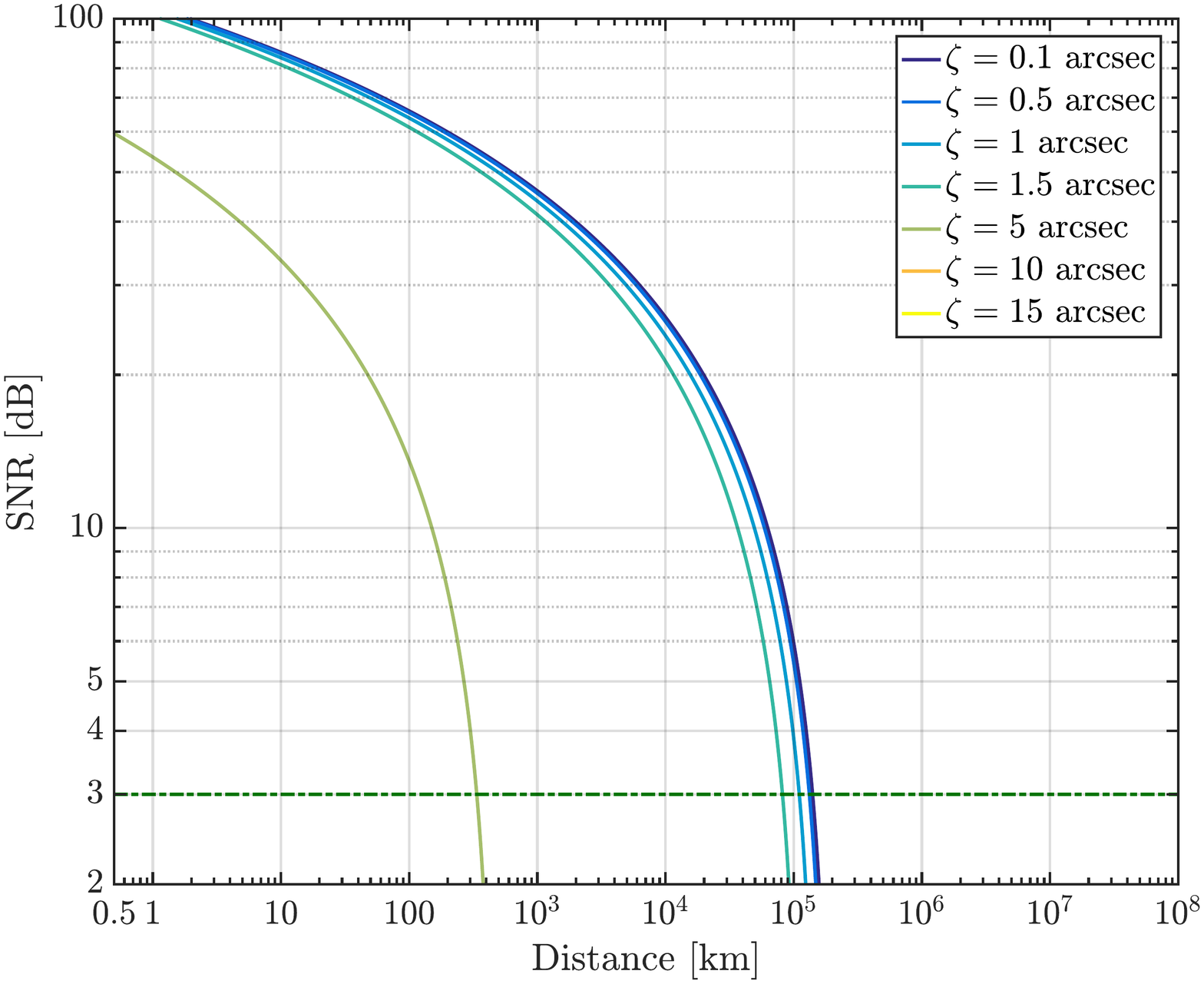}
	\caption{Signal to noise ratios ($\mathrm{SNR}$) as a function of interterminal distance and for varying attitude errors $\zeta$. Hardware parameters are those summarized in Table \ref{tab:sigBeamSumm}, and 
	the threshold for acquisition is $\mathrm{SNR}_{\mathrm{dB}}^* = 3$ (dash-dotted green line).}
	\label{fig:AcqComp_new}
\end{figure}

\begin{figure}[ht]
	\centering
		\includegraphics[width=1\textwidth]{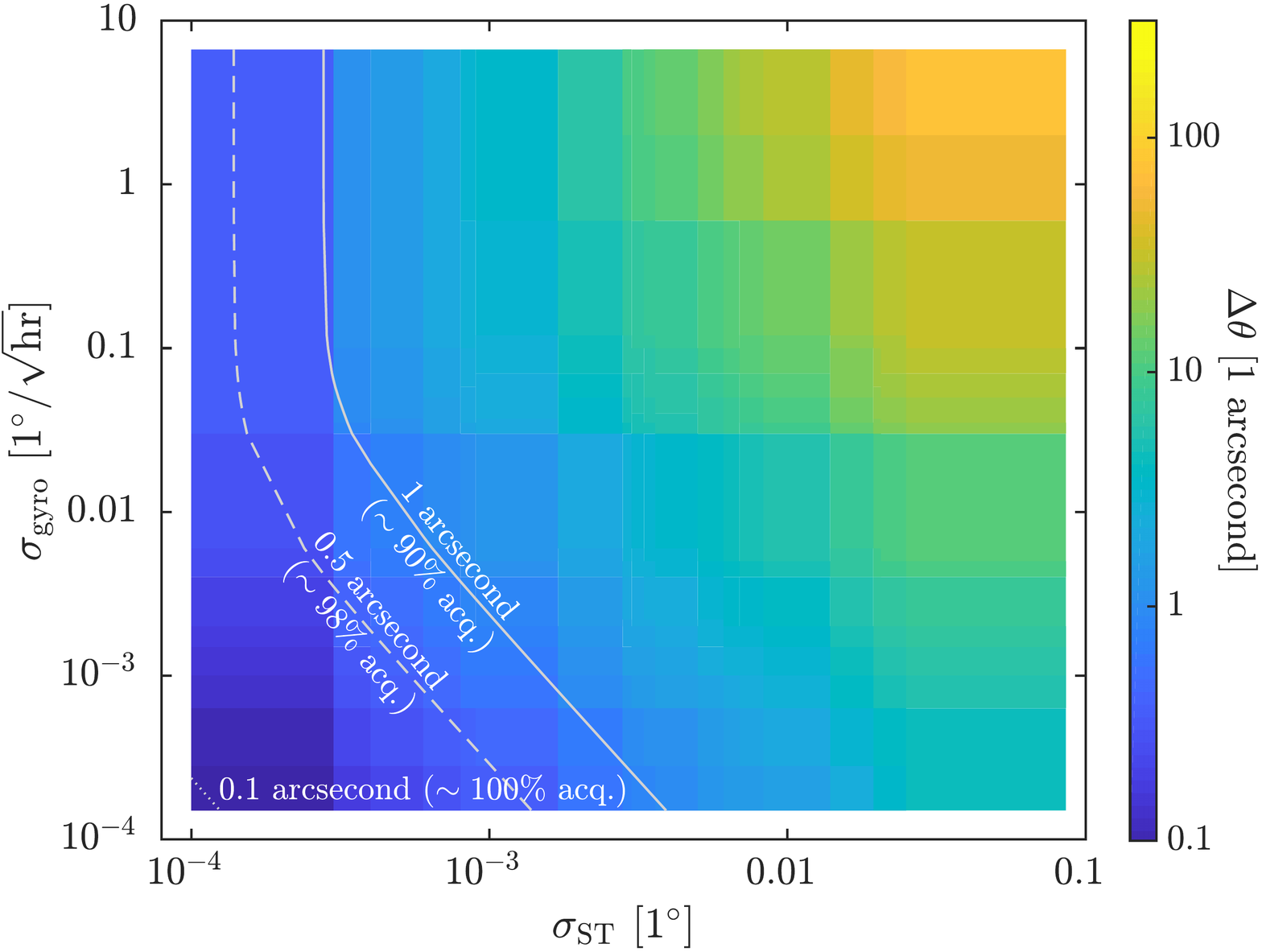}
	\caption{Attitude knowledge error as a function of gyroscope and star tracker performance. }
	\label{fig:AttKnowErr}
\end{figure}

\begin{figure}[ht]
	\centering
		\includegraphics[width=1\textwidth]{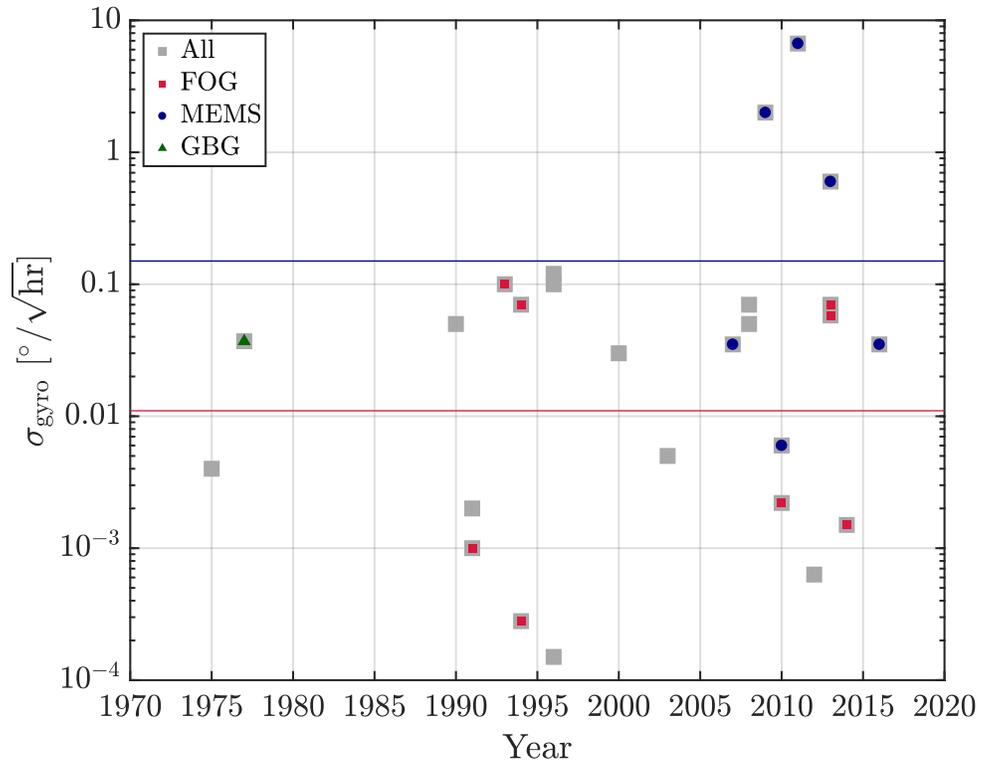}
	\caption{Gyroscope performance over time as a function of gyroscope type. Typical MEMS gyroscope performance (blue line) is about an order of magnitude worse than those for widely-used fiber optic gyroscopes. 
	On the other hand, MEMS gyros are more amendable to miniaturization and future performance capabilities may enable the requisite performance Fig. \ref{fig:MEMSGyro}). }
	\label{fig:Gyrocap}
\end{figure}

\begin{figure}[ht]
	\centering
		\includegraphics[width=1\textwidth]{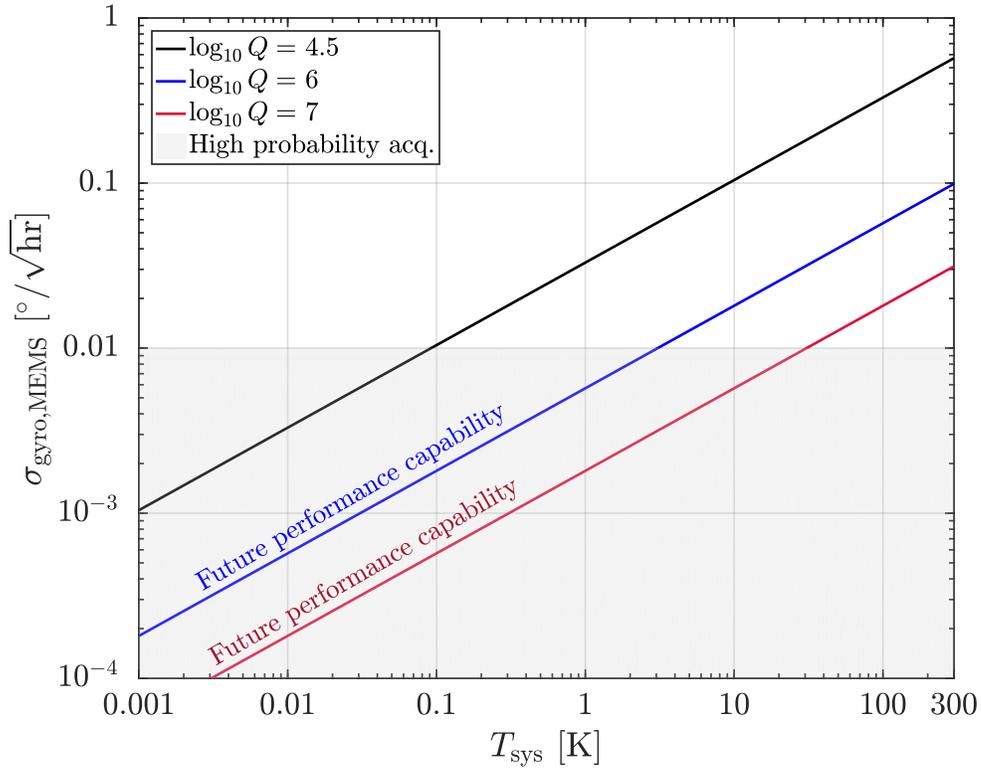}
	\caption{MEMS gyroscope performance as a function of quality factor $Q$ (a smaller value of which represents greater mechanical losses due to damping), and as a function of system temperature, reflecting 
	thermal noise. Lines representing current capabilities (black) and future capabilities based on a literature review (blue and red lines) are shown. For reasonable system temperatures, thermal noise-limited performance
	may be achievable in the near future.}
	\label{fig:MEMSGyro}
\end{figure}

\begin{figure}[ht]
	\centering
		\includegraphics[width=1\textwidth]{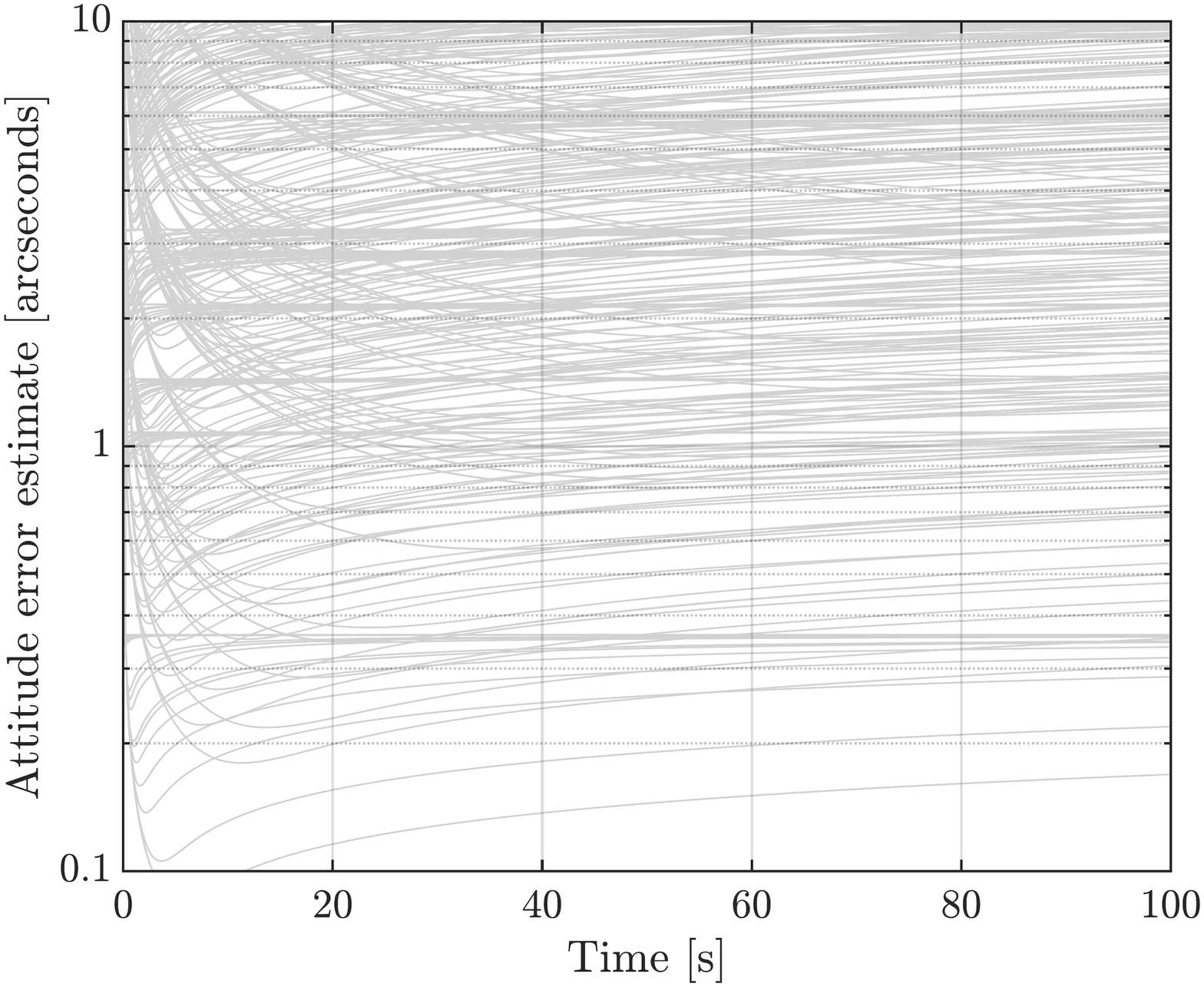}
	\caption{Least-squares Kalman filter models for all combinations of star tracker and gyroscope data. The minima from these iteratively-calculated curves \citep{hemmati2006} are used to generate Fig. \ref{fig:AttKnowErr}.}
	\label{fig:IndAttKnowErr}
\end{figure}

Is it possible to achieve the $\zeta \sim 0.01-1~\mathrm{arcsec}$ attitude errors that would enable interplanetary lasercom networks? Here we explore whether the contribution from attitude knowledge
can be decreased to sufficiently low values. We assume the attitude knowledge comes from star trackers and gyroscopes. Using a range of star tracker and gyroscope data and using an iterative Kalman filter
model \citep{hemmati2006}, in Fig. \ref{fig:AttKnowErr} we plot the contribution to $\zeta$ from star tracker and gyroscope performance. In Fig. \ref{fig:Gyrocap}, we show gyroscope performance over time as a function of gyroscope type
and in Fig. \ref{fig:MEMSGyro}, we show how physics-limited MEMS gyroscrope performance may evolve due to manufacturing capabilities. In this particular case, we show how MEMS gyroscope performance varies as a function of 
quality factor $Q$ (for which larger values represent smaller losses from dissipation) and system temperature (which leads to thermal noise; see \citet{zotov2014quality} and \citet{leland2005mechanical} for further details). Individual results from the Kalman filter model over the range of star tracker and 
gyroscope data points that lead to Fig. \ref{fig:AttKnowErr} are shown in Fig. \ref{fig:IndAttKnowErr}.

\section{Discussion and Conclusions}
\subsection{Summary and Discussion}
In this paper, we have considered several topics relevant to the development of interplanetary optical communication networks for small satellites. We have
\begin{itemize}
\item Constructed an analytical model for the mutual acquisition probability between two optical terminals,
\item Calculated the optimum beam width for acquisition over all angles,
\item Calculated the optimum beam width for a given off-pointing $\dvt$,
\item Constructed a notional interplanetary optical communication network and calculated its cost, and
\item Identified regimes of the parameter space whose future development would enable the implementation of interplanetary optical networks.
\end{itemize}
In recent years, picosats have seen increasing use for academic and commercial purposes, and their relatively low cost,
set form factor, and reliance on commercial/off-the-shelf hardware makes them attractive for further development, especially where large numbers of assets are involved. At the same time, however, these same qualities make their implementation difficult in circumstances where precision is involved. The purpose of this paper has been to highlight the quantities that are important to consider if picosats are to be implemented for optical communication networks at the interplanetary scale. Given the push to characterize the bodies of interplanetary space for both resource identification and security purposes, as well as the possibility of forming human habitats on these bodies or Mars, it seems worthwhile to further consider. 
\subsection{Recommendations and Conclusions}
In this paper, we have constructed a series of simple analytical models to aid us in characterizing optical terminals.
These models can be used to inform more sophisticated models, as well as to motivate the exploration of further technological development to enable interplanetary-scale optical networks---especially those in the challenging part of the design space occupied by the otherwise attractive picosat. Future work should explore optimizing interplanetary networks further. The geometry can be accounted for more precisely, and more economical constellations be explored. In particular, the ``resiliency'' and reliability of such networks from a data integrity and concept of operations point of view should be explored; one possible means is through graphical analysis. In this paper, we focused on MEMS gyroscopes as a potential enabling technology, but attitude control technologies for picosats should also be further explored. Future work should also explore other hardware options, and look more in depth at the technology frontiers (and potential disruptive technologies) that would enable interplanetary picosat networks. 

\newpage
\mbox{~}
\clearpage
\bibliographystyle{plainnat}
\bibliography{LaserComRefs}

\end{document}